\documentclass[aps,twocolumn,superscriptaddress,showpacs,pre]{revtex4}
\usepackage{graphicx}

\begin{document}
\title{Extreme nonlinear electrodynamics in metamaterials with very small linear dielectric permittivity}

\author{A. Ciattoni}
\affiliation{Consiglio Nazionale delle Ricerche, CNR-SPIN, 67100 L'Aquila, Italy \\ and Dipartimento di Fisica, Universit\`{a} dell'Aquila, 67100
L'Aquila, Italy}

\author{C. Rizza}
\affiliation{Dipartimento di Ingegneria Elettrica e dell'Informazione, Universit\`{a} dell'Aquila, 67100 Monteluco di Roio (L'Aquila), Italy}

\author{E. Palange}
\affiliation{Dipartimento di Ingegneria Elettrica e dell'Informazione, Universit\`{a} dell'Aquila, 67100 Monteluco di Roio (L'Aquila), Italy}

\date{\today}

\begin{abstract}
We consider a sub-wavelength periodic layered medium whose slabs are filled by arbitrary linear metamaterials and standard nonlinear Kerr media and
we show that the homogenized medium behaves as a Kerr medium whose parameters can assume values not available in standard materials. Exploiting such
a parameter availability, we focus on the situation where the linear relative dielectric permittivity is very small thus allowing the observation of
the extreme nonlinear regime where the nonlinear polarization is comparable with or even greater than the linear part of the overall dielectric
response. The behavior of the electromagnetic field in the extreme nonlinear regime is very peculiar and characterized by novel features as, for
example, the transverse power flow reversing. In order to probe the novel regime, we consider a class of fields (transverse magnetic nonlinear guided
waves) admitting full analytical description and we show that these waves are allowed to propagate even in media with $\epsilon<0$ and $\mu
>0$ since the nonlinear polarization produces a positive overall effective permittivity. The considered nonlinear waves exhibit, in addition to the
mentioned features, a number of interesting properties like hyper-focusing induced by the phase difference between the field components.
\end{abstract}
\pacs{78.67.Pt, 42.65.Tg}
\maketitle
\section{Introduction}
Electromagnetic propagation through metamaterials has stimulated, in the last decade, an intense research activity with two main purposes: the
identification of artificial structures exhibiting anomalous values of the permittivity $\epsilon$ and the permeability $\mu$ \cite{Ramak1,CaiSha}
and the investigation of the electromagnetic phenomenology resulting from the unusual electromagnetic properties. Leading examples of the novel class
of phenomena supported by metamaterials are superlensing \cite{Pendr1,FangLe}, optical cloaking \cite{Pendr2,Schuri} and photonic circuits
\cite{Enghet}. Recently, a good deal of attention has been devoted to the investigation of metamaterials characterized by a very small dielectric
permittivity ($\epsilon$-near-zero materials) since they host an electromagnetic regime where the magnetic field displays static features
\cite{Aluuuu} and the tunneling of electromagnetic energy through sub-wavelength channels has been proposed \cite{Silver} and experimentally observed
\cite{Liuuuu,Edward}. Metamaterials exhibiting remarkable nonlinear properties have been investigated as well \cite{Zharov} and various soliton
manifestations \cite{Shadri,Lazari,Zharoa,Hegdee} have been considered. The investigation of metamaterial nonlinear properties is particularly
important in that it can lead to overcoming one of the fundamental limit of nonlinear optics, the fact that most of the optical materials have a
relatively weak nonlinear response. The main idea is that the local electromagnetic fields of the inclusions in the metamaterial can be much larger
than the average value of the field thus producing an enhancement of the nonlinear response \cite{Pendr3,Shadr2,Obrien}. The problem of achieving a
substantial enhancement of the nonlinear response has also been considered within the more general subject of composite structures homogenization
\cite{Stroud,Neeves,Sipeee,Fische,Huangg}, and the strategy is always that of conceiving a microscopic inhomogeneous structure concentrating the
field within the nonlinear constituents.

The full exploitation of the nonlinear response is possible only if the nonlinear polarization is not a small perturbation to the linear part of the
electric displacement field and generally this is achieved through nonlinearity enhancement or by means of resonant processes or photorefractive
processes where the large nonlinearities come at the cost of a large time response. However, as shown in this paper, the interplay between the linear
part of the electric displacement field and the nonlinear polarization can be made efficient even by following the opposite route, i.e. by reducing
the linear polarization. We therefore devise a nonlinear medium with a very small linear dielectric constant since it is a natural setting for the
observation of the electromagnetic regime where the nonlinear response does not play the role of a mere perturbation.

In this paper we consider a periodic layered composite whose slabs are filled either with linear media with arbitrary permittivity and permeability
or by standard isotropic focusing or defocusing Kerr media. Exploiting a suitable extension of the well known technique generally used for describing
the homogenization of linear layered composites, we show that the homogenized medium is characterized by effective constitutive relations formally
coinciding with those of a standard Kerr medium. We note that the parameters characterizing such an effective response can be simply tailored through
a suitable choice of the composite underlying constituents and that, due the large freedom in choosing both constituent media and their volume
filling fraction, the design of linear and nonlinear properties can be independently performed. Therefore our composite medium allows a full and
efficient engineering of the Kerr nonlinear response. More interestingly, we prove that the effective response parameters span very wide ranges
encompassing values not available in standard media. The standard isotropic Kerr response (in the the frequency domain) generally depends on two
parameters usually denoted with $A$ and $B$ \cite{Boyddd} in term of which we define $\chi =A$ and $\gamma = B/2A$ and the available values of
$\gamma$ belong to the range $0<\gamma<3$ (depending on the actual mechanism supporting the Kerr response). In this paper we prove that the parameter
$\gamma$ appearing in the effective Kerr nonlinear response can, in principle, assume any value (encompassing negative and very large values) and we
discuss the impact of the exotic values of $\gamma$ on the electromagnetic phenomenology.

Confining our attention to transverse magnetic (TM) fields, we show that it is possible to design a nonlinear Kerr medium whose linear dielectric
permittivity is much smaller than one and therefore able to host the electromagnetic regime  where nonlinearity can not be regarded as a perturbation
(extreme nonlinear regime). As a first general electromagnetic effect characteristic of the extreme nonlinear regime, we discuss the transverse power
flow reversing, i.e. the fact that, for an electromagnetic beam, the Poynting vector on the beam propagation axis can be antiparallel to Poynting
vector on the beam lateral sides \cite{Ciattt}. Within the extreme nonlinear regime, the full exploitation of the nonlinear response can be achieved
and, in order to discuss the consequent novel electromagnetic phenomenology, we consider a class of nonlinear guided waves admitting full analytical
treatment. A number of the obtained nonlinear guided waves exhibit, as expected, transverse power flow reversing. We prove that, in the situation
where $\epsilon <0$, $\mu >0$ and $\chi>0$ (or in the equivalent case $\epsilon >0$, $\mu <0$ and $\chi<0$) these waves are allowed to propagate
through the medium if $-1 <\gamma <0$, in striking contrast to the fact that, in a linear medium with the same permittivity and permeability, any
wave would be evanescent. This is possible since nonlinearity overcomes the linear contribution to the electric displacement field producing an
effective nonlinear dielectric permittivity able to support propagating waves. In addition we identify a mechanism which, combining the extreme
nonlinear regime with the properties of the effective medium response with $\gamma \simeq -1$, produces an hyper-focusing of the considered nonlinear
guided waves. In the situation $\epsilon >0$, $\mu >0$ and $\chi<0$ (or in the equivalent case $\epsilon <0$, $\mu <0$ and $\chi>0$) the nonlinear
guided waves exist in the wider range $-\infty < \gamma < -1$ and $0 < \gamma < +\infty$ thus allowing to probe the Kerr nonlinear response for
large values of $|\gamma|$. Since the phase difference between the components of the considered waves is $\pi /2$, we predict that, for $\gamma
\rightarrow +\infty$, an extreme compensation occurs (within the effective nonlinear polarization) and supports a field whose transverse $E_x$ and
longitudinal $E_z$ components are such that $|E_x| \simeq |E_z|$, i.e they share the same profile.

The paper is organized as follows. In Sec.II we investigate the homogenization of a one-dimensional periodic layered medium comprising both linear
(metamaterials) and nonlinear (Kerr media) slabs and we derive the constitutive relations characterizing the effective medium. In Sec.III we focus on
the transverse magnetic field configuration and we discuss the effective nonlinear response engineering and the allowed extreme nonlinear regime,
together with the transverse power flow reversing effect. In Sec.IV we investigate a class of nonlinear guided waves belonging to the extreme
nonlinear regime and we discuss a number of their peculiar properties. The existence of the considered nonlinear guided waves is investigated in
Appendix A.
\section{Homogenization of a 1-D nonlinear layered medium}
Consider a monochromatic electromagnetic field (whose time dependence is $\exp(-i \omega t)$) propagating through a metamaterial layered medium
consisting of periodically repeating, along the $y-$axis, $N$ layers of different media of thicknesses $d_j$ ($j=1...N$), so that the structure
spatial period is $d= \sum_{i=j}^N d_j$ (see Fig.1 where the case $N=4$ is depicted). Each one of these $N$ media can be either a linear metameterial
with arbitrary dielectric and magnetic properties or a cubic standard nonlinear medium so that, the electromagnetic response of the $j-$th medium is
modelled by the general constitutive relations
\begin{eqnarray} \label{micr}
{\bf D}_j &=& \epsilon_0 \epsilon_j {\bf E}_j + \epsilon_0 \chi_j [({\bf E}_j \cdot {\bf E}_j^*) {\bf E}_j + \gamma_j ({\bf E}_j \cdot {\bf E}_j)
              {\bf E}_j^* ], \nonumber \\
{\bf B}_j &=& \mu_0 \mu_j {\bf H}_j,
\end{eqnarray}
\begin{figure}
\includegraphics[width=0.45\textwidth]{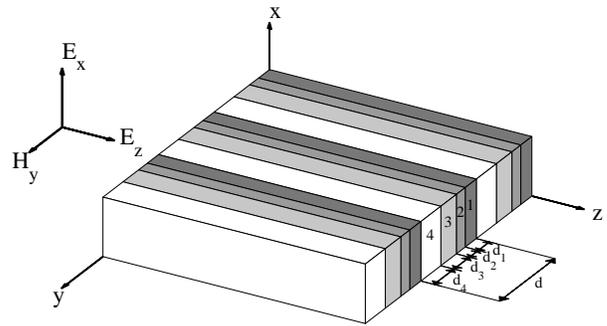}
\caption{Geometry of the nonlinear layered composite and TM electromagnetic field configuration.}
\end{figure}
where $\epsilon_0$ and $\mu_0$ are the vacuum permittivity and permeability constants, whereas ${\bf E}_j$, ${\bf D}_j$, ${\bf H}_j$, ${\bf B}_j$ are
the complex amplitude of the local electromagnetic field vectors, $\epsilon_j$ and $\mu_j$ are the relative permittivity and permeability of the
$j-$th layer, $\chi_j$ and $\gamma_j$ are the standard parameters characterizing the isotropic cubic nonlinear response of the $j-$th layer
\cite{Boyddd}. Evidently, $\chi_j = 0$ if the $j-$th layer is filled by a linear medium and $\mu_j =1$ for a nonlinear dielectric. The linear
dielectric constants $\epsilon_j$ are here regarded as arbitrary complex numbers since the layers can be filled with lossy and active media. If the
field vacuum wavelength is much greater than the spatial period ($\lambda = 2 \pi c / \omega \gg d$) the considered periodically nonlinear stratified
medium can be homogenized, i.e. its electromagnetic response can be shown to coincide with that of a suitable homogeneous medium. In order to obtain
the overall effective response, note that any component of the local electromagnetic field can be assumed, within each layer, independent on $y$
since the layers are extremely small ($d \ll \lambda$). Although this is a very reasonable physical assumption, it can be rigorously proven
exploiting the well known powerful two-scale expansion method \cite{Allair,Felbac}. Each of the physical observable (macroscopic) electromagnetic
field vectors, say $\bf V$ ($\bf V = E,D,B,H$), is obtained by averaging (along the $y-$ axis) the layer local fields over the period $d$, so that
\begin{equation}  \label{average}
{\bf V} \equiv \langle {\bf V}_j \rangle = \sum_{j=1}^{N} f_j {\bf V}_j
\end{equation}
where $f_j = d_j / d$ is the volume filling fraction of the $j-$th medium ($\sum_{j=1}^N f_j = 1$) and the averaging has been performed by exploiting
the uniformity along $y$ of the local fields ${\bf V}_j$. Within each unit cell (of thickness $d$), at each plane interface between the $j-$th and
$(j+1)-$th layer, the local fields have to satisfy the electromagnetic boundary conditions (continuity of the tangential component of electric and
magnetic fields and continuity of the normal components of the displacement and magnetic induction fields) so that the local fields are joined by the
relations
\begin{eqnarray} \label{boundary}
E_{jx} =& E_{(j+1)x}, \quad \quad H_{jx} &= H_{(j+1)x}, \nonumber \\
E_{jz} =& E_{(j+1)z}, \quad \quad H_{jz} &= H_{(j+1)z}, \nonumber \\
D_{jy} =& D_{(j+1)y}, \quad \quad B_{jy} &= B_{(j+1)y}
\end{eqnarray}
where $j=1...(N-1)$. Averaging the local fields $B_{jx}$, $B_{jz}$ and $H_{jy}$ and exploiting the second of Eqs.(\ref{micr}) together with magnetic
boundary conditions of Eqs.(\ref{boundary}), it is straightforward to prove that the macroscopic fields ${\bf B} = \langle {\bf B}_j \rangle$ and
${\bf H} = \langle {\bf H}_j \rangle$  are related by
\begin{equation} \label{BBB}
{\bf B} = \mu_0 \mu_{eff} {\bf H}
\end{equation}
where the relative magnetic permeability $\mu_{eff}$ is the diagonal tensor $\mu_{eff} = diag [\langle \mu_j \rangle, \langle \mu_j^{-1}
\rangle^{-1},\langle \mu_j \rangle]$, which is a very well-established result concerning the homogenization of layered media \cite{CabuzF}. As far as
the dielectric response is concerned, exploiting the first of Eqs.(\ref{micr}) together with the fact that $E_{jx} = \langle E_{jx} \rangle = E_x$,
$E_{jz} = \langle E_{jz} \rangle =E_z$ and $D_{jy} = \langle D_{jy} \rangle = D_y$ (obtained by combining the electric boundary conditions of
Eqs.(\ref{boundary}) and Eq.(\ref{average})) we obtain
\begin{eqnarray} \label{DD}
{\bf D}_\perp &=& \epsilon_0  \langle \epsilon_j \rangle {\bf E}_\perp  \nonumber \\
&+& \epsilon_0 \left[ \langle \chi_j \rangle ({\bf E}_\perp \cdot {\bf E}_\perp^*) {\bf E}_\perp + \langle \chi_j \gamma_j \rangle ({\bf E}_\perp
\cdot {\bf E}_\perp) {\bf E}_\perp^* \right]  \nonumber \\
&+& \epsilon_0 \left[ \langle \chi_j |E_{jy}|^2 \rangle {\bf E}_\perp + \langle \chi_j \gamma_j E_{jy}^2 \rangle {\bf E}_\perp^* \right], \nonumber \\
D_y &=& \epsilon_0 \epsilon_j E_{jy}  \nonumber \\
&+&\epsilon_0 \chi_j \left[ ({\bf E}_\perp \cdot {\bf E}_\perp^*) E_{jy} + \gamma_j ({\bf E}_\perp \cdot {\bf E}_\perp) E_{jy}^* \right]  \nonumber \\
&+& \epsilon_0 \chi_j \left[ (1+\gamma_j) |E_{jy}|^2 E_{jy}  \right]
\end{eqnarray}
where ${\bf D}_\perp = D_x \hat{\bf e}_x + D_z \hat{\bf e}_z$ and ${\bf E}_\perp = E_x \hat{\bf e}_x + E_z \hat{\bf e}_z$ are the transverse parts of
the macroscopic displacement and electric fields, respectively and $D_y$ is the $y-$component of the macroscopic displacement field. Note that in the
second of Eqs.(\ref{DD}) no averaging has been performed. As a consequence of the layers nonlinear dielectric behavior, the first of Eqs.(\ref{DD})
contains terms where the squares of $E_{jy}$ is suitably averaged. In order to derive the effective medium response, the second of Eqs.(\ref{DD}) has
to be solved to express $E_{jy}$ ($j=1...N$) as a function of the macroscopic fields $\bf E_\perp$ and $D_y$. This can be pertubatively done by
noting that, since we are considering the standard Kerr effect for each nonlinear layer \cite{Boyddd}, the terms containing $\chi_j$ are much smaller
than the term $\epsilon_0 \epsilon_j E_{jy}$ so that, up to the first order in the field cubic terms, from the second of Eqs.(\ref{DD}) we obtain
\begin{eqnarray} \label{EE}
\epsilon_0 E_{jy} &=& \frac{1}{\epsilon_j} D_y - \frac{\chi_j} {\epsilon_j^2} \left[({\bf E}_\perp \cdot {\bf E}_\perp^*) D_y + \gamma_j
      ({\bf E}_\perp \cdot {\bf E}_\perp) D_y^* \right] \nonumber \\
       &-& \chi_j \frac{1+\gamma_j}{\epsilon_0^2 \epsilon_j^4} |D_y|^2 D_y.
\end{eqnarray}
Averaging Eq.(\ref{EE}) we obtain a relation joining the macroscopic field $E_y$ and the fields ${\bf E}_\perp$ and $D_y$ so that this relation, if
its nonlinear contributions are much smaller than the leading linear term, can be inverted to perturbatively yield $D_y$ as a function of ${\bf
E}_\perp$ and $E_y$. Therefore, up to the first order in the field cubic terms, we obtain
\begin{eqnarray} \label{Dy}
D_y &=& \epsilon_0 \langle \epsilon_j^{-1} \rangle^{-1} E_y \nonumber \\
    &+& \epsilon_0 \left[ \frac{\langle \chi_j \epsilon_j^{-2} \rangle} {\langle \epsilon_j^{-1} \rangle^{2}} {\bf E}_\perp \cdot {\bf E}_\perp^*
        +\frac{\langle \chi_j (1+\gamma_j) \epsilon_j^{-4} \rangle}{2 \langle \epsilon_j^{-1} \rangle^{4}} |E_y|^2 \right] E_y \nonumber \\
    &+& \epsilon_0 \left[ \frac{\langle \chi_j \gamma_j \epsilon_j^{-2} \rangle} {\langle \epsilon_j^{-1} \rangle^{2}} {\bf E}_\perp \cdot {\bf E}_\perp
        +\frac{\langle \chi_j (1+\gamma_j) \epsilon_j^{-4} \rangle}{2 \langle \epsilon_j^{-1} \rangle^{4}} E_y^2 \right] E_y^*. \nonumber \\
\end{eqnarray}
Substituting the fields $E_{jy}$ of Eqs.(\ref{EE}) into the first of Eqs.(\ref{DD}) and using Eq.(\ref{Dy}) (neglecting everywhere the terms
containing powers of $\chi_j$ higher than one) we obtain
\begin{eqnarray} \label{Dp}
{\bf D}_\perp &=& \epsilon_0 \langle \epsilon_j \rangle {\bf E}_\perp \nonumber \\
    &+& \epsilon_0 \left[ \langle \chi_j \rangle {\bf E}_\perp \cdot {\bf E}_\perp^*
        +\frac{\langle \chi_j \epsilon_j^{-2} \rangle}{\langle \epsilon_j^{-1} \rangle^{2}} |E_y|^2 \right] {\bf E}_\perp \nonumber \\
    &+& \epsilon_0 \left[ \langle \chi_j \gamma_j \rangle {\bf E}_\perp \cdot {\bf E}_\perp
        +\frac{\langle \chi_j \gamma_j \epsilon_j^{-2} \rangle}{\langle \epsilon_j^{-1} \rangle^{2}} E_y^2 \right] {\bf E}_\perp^*.
\end{eqnarray}
Equations (\ref{Dy}) and (\ref{Dp}) solely contain the macroscopic fields so that they are the constitutive dielectric relations characterizing the
effective medium obtained by the homogenization of the considered 1D layered medium. The obtained dielectric response shows that the effective medium
behaves like an anisotropic Kerr medium whose nonlinear properties can be tailored by suitably choosing the underlying layered composite. Note that,
in the specific case of linear layers (i.e. $\chi_j =0$ for all $j$), Eqs. (\ref{Dy}) and (\ref{Dp}) yield
\begin{equation}
{\bf D}= \epsilon_0 \epsilon_{eff} {\bf E}
\end{equation}
where the relative dielectric permittivity $\epsilon_{eff}$ is the diagonal tensor $\epsilon_{eff} = diag [\langle \epsilon_j \rangle, \langle
\epsilon_j^{-1} \rangle^{-1},\langle \epsilon_j \rangle]$, reproducing a well known result concerning homogenization of linear layered media
\cite{Belovv,CabuzF}.
\section{Extreme electrodynamics of TM fields}
Electromagnetic propagation through the homogenized medium of Sec.II is described by the macroscopic Maxwell equations
\begin{eqnarray} \label{Maxwell}
\nabla \times {\bf E} &=& i \omega {\bf B}, \nonumber \\
\nabla \times {\bf H} &=& -i \omega {\bf D}
\end{eqnarray}
and the constitutive relations of Eqs.(\ref{BBB}), (\ref{Dy}) and (\ref{Dp}). Hereafter, we focus on Transverse Magnetic (TM) electromagnetic fields
(see Fig.1) of the form
\begin{eqnarray} \label{TMfield}
{\bf E} &=& E_x (x,z) \hat{\bf e}_x + E_z (x,z) \hat{\bf e}_z, \nonumber \\
{\bf H} &=& H_y (x,z) \hat{\bf e}_y
\end{eqnarray}
for which the constitutive relations become
\begin{eqnarray} \label{TMconst}
{\bf D} &=& \epsilon_0 \epsilon {\bf E} + \epsilon_0 \chi [({\bf E} \cdot {\bf E}^*) {\bf E} + \gamma({\bf E} \cdot {\bf E}) {\bf E}^* ], \nonumber \\
{\bf B} &=& \mu_0 \mu {\bf H},
\end{eqnarray}
where
\begin{eqnarray} \label{parameters}
\epsilon = & \displaystyle \sum_{j=1}^N f_j \epsilon_j, \quad \quad  \mu &= \left[\sum_{j=1}^N f_i \mu_j^{-1} \right]^{-1}, \nonumber \\
\chi = & \displaystyle \sum_{j=1}^N f_j \chi_j, \quad \quad \gamma &= \frac{1}{\chi} \sum_{j=1}^N f_j \chi_j \gamma_j.
\end{eqnarray}
Therefore the effective dielectric response experienced by a TM field coincides with the standard isotropic Kerr response. In analogy to what happens
in linear layered media we note, from Eqs.(\ref{parameters}), that the permittivity $\epsilon$ and the permeability $\mu$ are the weighted and
harmonic weighted means, respectively, so that the former is bounded by the minimum and maximum of its microscopic values whereas the latter is
unbounded since some $\mu_j$ can be negative. It is worth stressing that the averaging of the microscopic linear parameters additionally allows an
efficient loss management since, combining lossy and gain media, one can design an effective medium whose effective parameters $\epsilon$ and $\mu$
have negligible imaginary parts \cite{Ramak2}. On the other hand, $\chi$ is the weighted mean of the constituents nonlinear Kerr coefficients and
this is in agreement with the results of Ref.\cite{Boydds}. More interestingly, $\gamma$ is generally not the weighted mean of its microscopic values
since some $\chi_j$ can be negative (corresponding to defocusing nonlinear layers), so that $\gamma$ can assume any positive and negative value. This
result is particularly interesting since, for standard cubic isotropic materials, there is, in general, a small number of available $\gamma$
(depending on the physical mechanism supporting the nonlinear response \cite{Boyddd}) whereas, for composite materials, it has been shown in
Ref.\cite{Sipeee} that $\gamma$ can span the whole range $0<\gamma<3$.

Even though Eqs.(\ref{TMconst}) formally coincides with the standard isotropic Kerr response, the considered layered material can support a nonlinear
electromagnetic phenomenology very different from that observed in a standard Kerr medium, as discussed in the following three subsections.

\subsection{Extreme nonlinear regime} \label{ENR}
Consider a layered medium for which $\epsilon$ is such that $0 < Re(\epsilon) \ll 1$, $|Im(\epsilon)| \ll Re(\epsilon)$ and the effective nonlinear
coefficient $\chi$ is of the same order of magnitude of underlying coefficients $\chi_j$ (i.e. $|\chi| \sim |\chi_j|)$. An effective dielectric
permittivity with a very small real part and a negligible imaginary part can be achieved by combining positive and negative standard dielectric
layers ($|Re (\epsilon_j)| > 1$) together with gain media (generally unavoidable since absorption due to negative dielectrics can not be in principle
neglected \cite{Ramak2}). For the sake of simplicity let us consider the case $\gamma =0$. If the electromagnetic field propagating through the
medium is such that
\begin{equation} \label{extr}
|E_x|^2 + |E_z|^2  \sim \left| \frac{\epsilon}{\chi} \right|,
\end{equation}
we have $|\chi_j ({\bf E} \cdot {\bf E}^*) | \sim |\epsilon| \ll |\epsilon_j|$ so that, in the first of Eqs.(\ref{micr}), the nonlinear part is much
smaller than the linear contribution $\epsilon_0 \epsilon_j {\bf E}_j$. Therefore, the nonlinear layers are in the presence of a field whose
intensity is sufficiently small for their response to be purely cubic and, as a consequence, the microscopic responses of Eqs.(\ref{micr}) hold and
the macroscopic constitutive relation in the first of Eqs.(\ref{TMconst}), holds as well. On the other hand, combining Eq.(\ref{extr}) and the first
of Eq.(\ref{TMconst}) we conclude that, as opposed to what happens in standard nonlinear materials, the considered layered medium is able to support
the extreme nonlinear regime where the electromagnetic field is such that the linear contribution $\epsilon {\bf E}$ and the nonlinear term $\chi
({\bf E} \cdot {\bf E}^*) {\bf E}$ in the overall dielectric response have the same order of magnitude. It is evident that the same argument, with
some slight changes, allows to prove that the extreme nonlinear regime is observable for any value of $\gamma$. Even though the advantages of such an
extreme nonlinear regime are self-evident, it is worth to compare it to the standard paraxial nonlinear optical situation where the overall
refractive index is $n_0 + \delta n$ where $n_0$ is the background linear refractive index and $\delta n = n_2 I << n_0$ is the Kerr nonlinear term
(where $I$ is the optical intensity). In this situation the field satisfies the equation
\begin{equation} \label{helmo}
\nabla ^2 {\bf E} + k_0^2(n_0^2 + 2 n_0 \delta n)  {\bf E} =0
\end{equation}
($k_0 = \omega / c$) and, within the paraxial regime, the field is of the form ${\bf E} = e^{ikz} {\bf A}$. At this point one chooses $k=k_0 n_0$
since, substituting the paraxial field into Eq.(\ref{helmo}) the term $k_0^2 n_0^2 {\bf E}$ (responsible for the fast spatial variation of the field)
is removed and the paraxial equation $i \frac{\partial {\bf A}}{\partial z} + \frac{1}{2k} \nabla^2_\perp  {\bf A}= -\frac{k}{n_0} \delta n {\bf A}$
is readily obtained by neglecting the term containing $\partial_z^2 {\bf A}$. Therefore, in standard paraxial nonlinear optics, $\delta n$ can play
an important role in Eq.(\ref{helmo}) since the large linear contribution proportional to $n_0 ^2$ is suppressed by the presence of the carrier plane
wave. In other words, the nonlinear Kerr contribution $\epsilon_0 \chi |{\bf E}|^2 {\bf E}$ is not requested to compete against the whole linear part
$\epsilon_0 \epsilon {\bf E}$ of the dielectric response. As opposed, in the case of the extreme nonlinear regime discussed in this section, this
competition can happen and therefore the nonlinearity is not confined to solely drive the slowly varying amplitude ${\bf A}$.
\subsection{Transverse power flow reversing} \label{PFR}
The extreme nonlinear regime can support propagation of beams characterized by exotic properties such as the reversing of the electromagnetic power
flow along the beam transverse profile \cite{Ciattt}. In order to discuss this phenomenon, consider a TM field describing a beam mainly propagating
along the $z$-axis, i.e. a field of the form $H_y (x,z) = e^{i K z} A(x,z)$ with the requirement $|\partial _z A| \ll K |A|$ ($K$ being a wave-vector
depending on the actual electromagnetic configuration). In this case Maxwell equations of Eqs.(\ref{Maxwell}) yield
\begin{eqnarray} \label{Hy}
H_y &=& \frac{\omega}{K} D_x, \nonumber \\
H_y &=& \frac{1}{\omega \mu_0 \mu} \left( K E_x + i \frac{\partial E_z}{\partial x} \right),
\end{eqnarray}
where it has been assumed that $\partial_z H_y \simeq iK H_y$ and $\partial_z E_x \simeq iK E_x$. Exploiting Eqs.(\ref{Hy}), the $z$-component of the
time averaged Poynting vector $S_z =(1/2) Re (H_y E_x^*)$ can be expressed through the equivalent relations
\begin{eqnarray} \label{Sz}
S_z &=& \frac{\omega}{2 K} Re (D_x E_x^*) \nonumber \\
S_z &=& \frac{1}{2\omega \mu_0 \mu} Re \left[ \left( K E_x + i \frac{\partial E_z}{\partial x} \right) E_x^* \right].
\end{eqnarray}
If, for example, $\epsilon>0$, $\mu >0$, $\chi<0$ and $\gamma = 0$, for an electromagnetic beam whose peak electric field strength is greater than
$\sqrt{|\epsilon / \chi|}$ (extreme nonlinear regime), it is evident from the first of Eqs.(\ref{TMconst}) that $D_x$ and $E_x$ are antiparallel
around the propagation axis (i.e. where $|\chi (\bf{E} \cdot \bf{E}^*)| > \epsilon$) and parallel elsewhere so that, from the first of
Eqs.(\ref{Sz}), $S_z$ is negative near the beam axis and positive elsewhere. In other words the beam is characterized by a power flow whose direction
reverses its sign along the transverse profile and this is due to sign flipping of $D_x$ along the wave transverse profile while $E_x$ does not
change its sign. Note that the sign flipping of $D_x$ corresponds to a sign flipping of $H_y$ (see the first of Eqs.(\ref{Hy})) so that, considering
the second of Eqs.(\ref{Hy}), this can happen without sign flipping of $E_x$ only if $\partial E_z / \partial x$ is not negligible with respect $K
E_x$. Therefore power flow reversing can take place only if the field has a transverse size comparable with $1/K$. It is worth stressing that the
discussed power flow reversing is very different from the effect that, in left handed metamaterials, the Poynting vector is antiparallel to the
carrier wave vector which is a consequence of the fact that, in such media, $\epsilon < 0$ and $\mu <0$ (with $n<0$). On the other hand, in our case,
$\mu > 0$ and the sign of the power flow is not uniform being controlled by the field intensity through the nonlinearity.
\subsection{Linear and nonlinear parameters design} \label{design}
In order to discuss the impact of the wide ranges of the parameters of Eqs.(\ref{parameters}), made possible by linear and nonlinear design, on
electromagnetic phenomenology, we consider the situation where the effective dielectric permittivity $\epsilon$ and magnetic permeability $\mu$ are
real, so that, after substituting the expression for the TM field of Eqs.(\ref{TMfield}) into Eqs.(\ref{Maxwell}), eliminating the magnetic field and
using Eqs.(\ref{TMconst}) we get
\begin{eqnarray} \label{system}
&& \frac{\partial ^2 U_z}{\partial \xi \partial \zeta} - \frac{\partial ^2 U_x}{\partial \zeta^2}  \nonumber \\
&& = \sigma_\epsilon \sigma_\mu U_x +  \sigma_\mu \left[ (1+\gamma) |U_x|^2 U_x + \left( U_x U_z^* + \gamma U_x^* U_z \right) U_z \right], \nonumber \\
&& \frac{\partial ^2 U_x}{\partial \xi \partial \zeta} - \frac{\partial ^2 U_z}{\partial \xi^2}  \nonumber \\
&& = \sigma_\epsilon \sigma_\mu U_z  + \sigma_\mu \left[ (1+\gamma) |U_z|^2 U_z +  \left( U_z U_x^* + \gamma U_z^* U_x \right) U_x \right], \nonumber \\
\end{eqnarray}
where dimensionless variables and fields have been introduced according to
\begin{eqnarray} \label{dimension}
x &= \displaystyle \frac{1}{\sqrt{|\epsilon \mu|} k_0} \xi, \quad \quad z =& \frac{1}{\sqrt{|\epsilon \mu|} k_0} \zeta, \nonumber \\
U_x &= \displaystyle \sqrt{\left|\frac{\chi}{\epsilon}\right|} E_x, \quad \quad U_z =& \sqrt{\left|\frac{\chi}{\epsilon}\right|} E_z,
\end{eqnarray}
and $k_0 = \omega /c$ whereas $\sigma _ \epsilon = \textrm{sign} (\epsilon \chi)$ and $\sigma_\mu = \textrm{sign} (\mu \chi)$ are the signs of the
products $\epsilon \chi$ and $\mu \chi$, respectively. As opposed to the linear regime where electric field behavior solely depends on the sign of
$\epsilon \mu$ \cite{Ramak1}, from Eqs.(\ref{system}) we note that, in the present approach, the nonlinear dynamics separately depends on the signs
of $\epsilon \chi$ and $\mu \chi$, i.e. the presence of the nonlinearity breaks the symmetry between the roles played by the signs of $\epsilon$ and
$\mu$ \cite{Scalor,Huwenz}. From Eqs.(\ref{dimension}) we note that $\epsilon$, $\mu$ and $\chi$ scale the actual physical size and amplitude of the
field so that, since the effective parameters can in principle be independently chosen (see Eqs.(\ref{parameters})), for each solution of
Eqs.(\ref{system}), a suitable layered medium can be designed in such a way that the actual electromagnetic field has prescribed geometrical size and
intensity (see Eqs.(\ref{dimension})). As an example, in the case of beam propagation, such an electromagnetic scaling freedom can allow to observe
nonparaxial feature of a beam whose transverse width is much greater than the vacuum wavelength (if $|\epsilon \mu|<<1$) or, on the contrary, to
observe the standard paraxial phenomenology for beams whose transverse width is much smaller than the vacuum wavelength (if $|\epsilon \mu|>>1$) and,
remarkably, this can be done by avoiding any unfeasible requirement on the intensity.

The parameter $\gamma$ plays a role fundamentally different since it cannot be generally removed from Eqs.(\ref{system}) with a field transformation.
Setting $U_x = A_x e^{i\phi_x}$ and $U_z = A_z  e^{i \phi_z}$ (where $A_x$, $A_z$, $\phi_x$ and $\phi_z$ are real), the nonlinear terms of
Eqs.(\ref{system}), namely $N_x = (1+\gamma) |U_x|^2 U_x + \left( U_x U_z^* + \gamma U_x^* U_z \right) U_z$ and $N_z = (1+\gamma) |U_z|^2 U_z +
\left( U_z U_x^* + \gamma U_z^* U_x \right) U_x $, can be written as
\begin{eqnarray} \label{nonli}
N_x &=& \left[ (1+\gamma) A_x^2 + \left(1+\gamma e^{-2i(\phi_x-\phi_z)}\right) A_z^2\right] A_x e^{i \phi_x}, \nonumber \\
N_z &=& \left[ \left(1+\gamma e^{2i(\phi_x-\phi_z)}\right) A_x^2 + (1+\gamma) A_z^2   \right] A_z e^{i \phi_z}, \nonumber \\
\end{eqnarray}
from which we note that $\gamma$ is responsible for a nonlinear coupling between the two field components which is highly sensitive to the phase
difference $\phi_x-\phi_z$. The case $\gamma = -1$ is particularly interesting since in Eqs.(\ref{nonli}) the terms proportional to $(1+\gamma)$
vanish so that each field component is nonlinearly driven only by the other component and such coupling is fully sensitive the phase difference. Also
intriguing is the regime $|\gamma|>>1$ since, two major opposite situations exist. If the phases of $U_x$ and $U_z$ are equal, from Eqs.(\ref{nonli})
it is evident that the overall nonlinear polarization is proportional to $(1+\gamma)$ so that, if $|\gamma| >>1$, the effective nonlinear response
can be enhanced. If, on the other hand, the phase difference between $U_x$ and $U_z$ is $\pi /2$, the overall nonlinear response of Eqs.(\ref{nonli})
solely contains the terms $(A_x^2 + A_z^2)$ and $\gamma (A_x^2-A_z^2)$, so that, if $|\gamma|>>1$ (for realistic bounded fields) the compensation
$A_x^2 \sim A_z^2$ has to occur in order to prevent the divergence of the term proportional to $\gamma$ in Eqs.(\ref{system}).
\section{Nonlinear guided waves}
As explained in the above section, the nonlinear response of the proposed Kerr metamaterial is easy to manage and, since the accesible ranges of its
parameters are very wide, one can devise situations where the medium supports a genuinely novel nonlinear electrodynamical phenomenology. In order to
probe the novel regime, we consider here a class of fields which are sufficiently simple to admit full analitical treatment and, at the same time,
rigged with enough structure to show many of the novel effects. More specifically we consider nonlinear guided waves propagating along the $z-$axis
of the form
\begin{eqnarray} \label{solit}
U_x(\xi,\zeta) &=& e^{i\beta \zeta} u_x(\xi), \nonumber \\
U_z(\xi,\zeta) &=& e^{i\beta \zeta} i u_z(\xi),
\end{eqnarray}
where $\beta$ is a real constant and the amplitudes $u_x$ and $u_z$ are real. Substituting the field of Eqs.(\ref{solit}) into Eqs.(\ref{system}) we
obtain
\begin{eqnarray} \label{syste}
&& -\beta \frac{d u_z}{d \xi} + \beta^2 u_x = \sigma_\mu \left[ \sigma_\epsilon  + \Psi_x \right] u_x, \nonumber \\
&& \beta \frac{d u_x}{d \xi} - \frac{d^2 u_z}{d \xi^2} = \sigma_\mu \left[ \sigma_\epsilon + \Psi_z \right] u_z
\end{eqnarray}
where
\begin{eqnarray} \label{Ga1}
\Psi_x (\xi) &=& (1+\gamma) u_{x}^2(\xi) + (1-\gamma) u_{z}^2 (\xi), \nonumber \\
\Psi_z (\xi) &=& (1-\gamma) u_{x}^2(\xi) + (1+\gamma) u_{z}^2 (\xi).
\end{eqnarray}
Using Eqs.(\ref{Ga1}), it is worth noting that, from the first of Eqs.(\ref{TMconst}), the vector $\bf D$ can be expressed as
\begin{eqnarray} \label{Dx}
D_x &=& \epsilon_0 \chi \sqrt{\left|\frac{\epsilon}{\chi}\right|^3} \left( \sigma_\epsilon + \Psi_x \right)   u_x e^{i \beta \zeta}, \nonumber \\
D_z &=& \epsilon_0 \chi \sqrt{\left|\frac{\epsilon}{\chi}\right|^3} \left( \sigma_\epsilon + \Psi_z \right) i u_z e^{i \beta \zeta}
\end{eqnarray}
from which we note that
\begin{eqnarray} \label{eeff}
\epsilon_{x}^{(NL)} &=& |\epsilon| \textrm{sign}(\chi) \left(\sigma_\epsilon + \Psi_x \right) \nonumber \\
\epsilon_{z}^{(NL)} &=& |\epsilon| \textrm{sign}(\chi) \left( \sigma_\epsilon + \Psi_z \right)
\end{eqnarray}
act, for the fields of Eqs.(\ref{solit}), as effective nonlinear dielectric permittivity. The quantities of Eqs.(\ref{Ga1}) play a fundamental role
in our discussion since it is evident that, if the conditions $|\Psi_x|<<1$ and $|\Psi_z|<<1$ do not hold along the profile of a nonlinear guided
wave, the linear and nonlinear contribution in Eq.(\ref{Dx}) are comparable, i.e. the considered nonlinear guided wave belongs to the extreme
nonlinear regime we have discussed in Sec.\ref{ENR}.

We consider solutions of Eqs.(\ref{syste}) with definite parity where $u_x$ and $u_z$ are spatially even ($u_x(\xi) = u_x(-\xi)$) and odd ($u_z(\xi)
= - u_z(-\xi)$), respectively and, as a consequence, we adopt the boundary conditions
\begin{equation} \label{boundary1}
\begin{array}{cc}
u_x(0) = u_{x0}, & u_z(0) = 0, \\
u_x(+\infty) = u_{x\infty}, &  u_z(+\infty) = u_{z\infty}.
\end{array}
\end{equation}
It is worth stressing that, due to the feasible possibility of arbitrary choosing the effective linear and nonlinear parameters characterizing the
effective nonlinear medium, we consider here solutions of Eqs.(\ref{syste}) and (\ref{boundary1}) for $\sigma_\epsilon = \pm 1$, $\sigma_\mu = \pm 1$
and, remarkably, for any real $\gamma$. In Appendix A we show that the system of Eqs.(\ref{syste}) is integrable and we derive the existence
conditions characterizing the nonlinear guided waves satisfying Eqs.(\ref{boundary1}), i.e, for each possible combinations of $\sigma_\epsilon$ and
$\sigma_\mu$, we derive a $\gamma$ dependent range of $u_{x \infty}^2$ (i.e. $u_{min}^2(\gamma) < u_{x \infty}^2 < u_{max}^2(\gamma)$) corresponding
to nonlinear waves existence.
\begin{table}
\caption{Guided waves existence ranges of $u_{x \infty}^2$ depending on $\gamma$, $\sigma_\epsilon$ and $\sigma_\mu$.}
\begin{ruledtabular}
\begin{tabular}{|c|c|c|}
             & $\sigma_\epsilon = -1 \quad \sigma_\mu=1$  &  $\sigma_\epsilon = -1 \quad \sigma_\mu=-1$     \\
\hline
$\gamma > 0$ &  & $\frac{\gamma}{\gamma^2 + 4 \gamma +1}< u_{x_\infty}^2 < \frac{1}{2} $ \\
\hline
$-1 < \gamma < 0$ & $u_{x_\infty}^2 < \frac{1}{2}$ &   \\
\hline
$\gamma < -1$ &  & $\frac{1}{1-\gamma} < u_{x_\infty}^2 < \frac{1}{2}$ \\
\end{tabular}
\end{ruledtabular}
\end{table}
The resulting phenomenology is reported in Table I. For each obtained nonlinear guided wave the propagation constant $\beta$ and the asymptotical
value $u_{z \infty}$ are given by (see Appendix A)
\begin{eqnarray} \label{beta}
\beta^2 &=& \frac{2 \gamma \sigma_\mu }{1+\gamma} (\sigma_\epsilon  + 2  u_{x \infty}^2) =
       \sigma_\mu \left[\sigma_\epsilon+\Psi_x(+\infty)\right], \nonumber \\
u_{z \infty}^2 &=& \frac{-\sigma_\epsilon - (1-\gamma) u_{x \infty}^2}{1+\gamma}
\end{eqnarray}
where the second expression for $\beta$ is obtained from the first by exploiting the first of Eqs.(\ref{Ga1}). It is worth noting that, for each
$u_{x \infty}$, these values of $\beta$ and $u_{z \infty}$ are obtained by requiring that the nonlinear guided wave is asymptotically spatially
uniform (i.e. by annulling the derivatives in Eqs.(\ref{syste}) for $\xi \rightarrow \infty$) and, exploiting Eqs.(\ref{eeff}), this implies that
\begin{eqnarray} \label{efflim}
\epsilon_{x}^{(NL)} (+\infty) &=& \frac{|\epsilon \mu|}{\mu} \beta^2, \nonumber \\
\epsilon_{z}^{(NL)} (+\infty) &=& 0.
\end{eqnarray}
From a physical point of view, the second of Eqs.(\ref{efflim}) states that the considered nonlinear waves can propagate through the medium only if,
asymptotically, the nonlinear term exactly balances the linear one to yield an overall vanishing dielectric constant, i.e.
\begin{equation}
\Psi_z(+\infty) = -\sigma_\epsilon.
\end{equation}
This proves that the nonlinear guided waves we are considering always belong to the extreme nonlinear regime, i.e. they can not be observed in
standard Kerr media. It is worth stressing that the sub-family of nonlinear guided waves with $u_{z \infty}=0$ do not require such an asymptotical
compensation mechanisms since, for $\xi \rightarrow \infty$, the second of Eqs.(\ref{syste}) vanishes together with $u_{z \infty}$. To sum up, an
asymptotically spatially uniform wave can exist only if $D_z(\infty)=0$ (since $\partial_x H_y = -i \omega D_z$) and this can be achieved by
requiring either $u_{z \infty}=0$ or $\epsilon_{z}^{(NL)} (+\infty) = 0$, the second situation being investigated in the present paper.

Since the waves we are investigating always belong to the extreme nonlinear regime, power flow reversing discussed in Sec.\ref{PFR} is expected. For
the field of Eqs.(\ref{solit}), the time averaged Poynting vector ${\bf S} = (1/2) Re [{\bf E} \times {\bf H}^*]$ is given by
\begin{eqnarray} \label{SSz1}
{\bf S} &=& S_0 \textrm{sign}(\mu) \left[ \left( \beta u_x - \frac{d u_z}{d \xi} \right) u_x \right] \hat{\bf e}_z, \nonumber \\
{\bf S} &=& S_0 \frac{\textrm{sign}(\chi)}{\beta} \left(\sigma_\epsilon + \Psi_x \right) u_x^2 \hat{\bf e}_z
\end{eqnarray}
where $S_0 = \sqrt{(\epsilon_0 |\epsilon|^3 )/(4 \mu_0 |\mu| |\chi|^2)}$ and the second expression is obtained from the first one by exploiting the
first of Eqs.(\ref{syste}) to eliminate the derivative of $u_z$ and the the first of Eqs.(\ref{Ga1}). Note that Eqs.(\ref{SSz1}) coincide with
Eqs.(\ref{Sz}) (with the identification $K=\sqrt{|\epsilon \mu|}k_0 \beta$), the difference lying in the fact that, for the nonlinear guided waves we
are considering, Eqs.(\ref{SSz1}) are exact due to the waves propagation invariance.
\subsection{Nonlinear guided waves for $\sigma_\epsilon = -1$ and $\sigma_\mu = 1$} \label{NGW1}
The situation $\sigma_\epsilon = -1$ and $\sigma_\mu = 1$ is very intriguing since $\textrm{sign}(\epsilon \mu) = \sigma_\epsilon \sigma_\mu = -1$
and therefore, in the absence of nonlinearity, no propagation can occur since the medium can support only evanescent waves. On the contrary, we have
shown that, in this situation (see first column of Table I), the nonlinear Kerr metamaterial can support propagating  nonlinear guided waves (i.e.
with real propagation constant $\beta$). In order to grasp this fact, we note, from the first of Eqs.(\ref{beta}), for $\sigma_\epsilon = -1$ and
$\sigma_\mu =1$, we obtain
\begin{equation} \label{betaq}
\beta^2 = -1 + \Psi_x(+\infty)
\end{equation}
from which it is evident that, in the absence of the nonlinearity, $\beta^2 =-1$, i.e. only evanescent waves exist. However, combining the first of
Eqs.(\ref{Ga1}) and the second of Eqs.(\ref{beta}), we obtain
\begin{equation}
\Psi_x (+\infty) = 1 -\frac{2 \gamma}{1+ \gamma} \left(1 - 2 u_{x \infty}^2 \right)
\end{equation}
from which it is evident that, for $-1 < \gamma <0$ and $u_{x_\infty} ^2<1/2$ (see the first column of Table I), $\Psi_x(+\infty)
> 1$ so that $\beta^2 >0$ and the nonlinear wave can propagate through the medium. From a physical point of view, the same result can be understood
by regarding the nonlinear guided wave as a background infinite plane wave with a distortion around $\xi=0$. The background nonlinear plane wave can
propagate through the medium since its amplitude is such that $\Psi_x(+\infty)>1$ so that the effective nonlinear dielectric permittivity in the
first of Eqs.(\ref{eeff}) is such that $\mu \epsilon_{x}^{(NL)} = |\epsilon \mu| \left[-1 +\Psi_x(+\infty)\right] > 0$. In other words, in the
extreme nonlinear regime, the nonlinear polarizability overcomes the linear dielectric contribution in such way that the sign of the effective
overall dielectric permittivity is opposed to that of the linear dielectric permittivity and the waves are consistently not evanescent.

\begin{figure}
\includegraphics[width=0.45\textwidth]{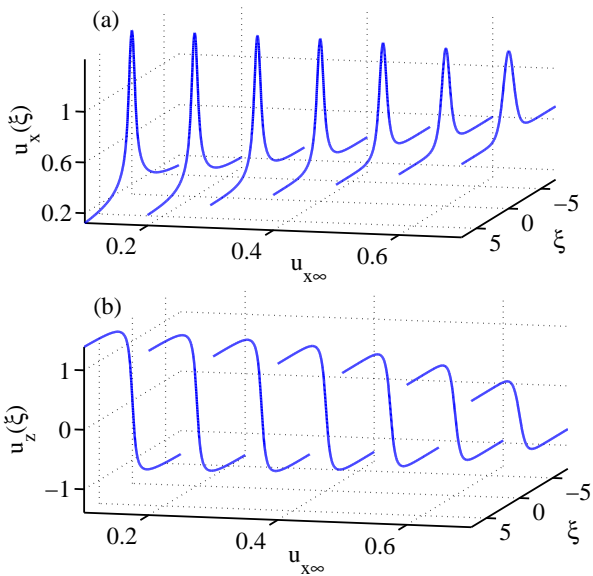}
\caption{Nonlinear guided waves transverse profile of $u_x(\xi)$ (subplot (a)) and $u_z(\xi)$ (subplot (b)) at different values of $u_{x \infty}$ for
$\sigma_{\epsilon}=-1$, $\sigma_{\mu}=1$, $\gamma=-1/2$.}
\end{figure}
\begin{figure}
\includegraphics[width=0.45\textwidth]{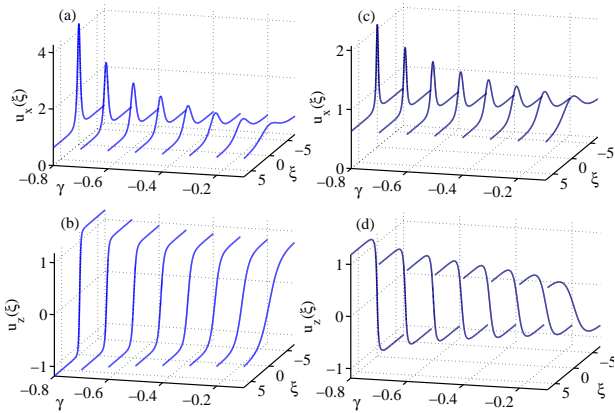}
\caption{Profiles of $u_x$ and $u_z$ of various nonlinear guided waves with $\sigma_\epsilon=-1$ and $\sigma_\mu=1$ and with $\gamma$ spanning the
range $-0.8<\gamma<-0.1$. All the waves are characterized by the same asymptotical value $u_{x\infty} =\sqrt{0.4}$. For each $\gamma$ two waves
exist, the first being reported in panel (a) and (b) and the second in panel (c) and (d)}
\end{figure}
\begin{figure}
\includegraphics[width=0.45\textwidth]{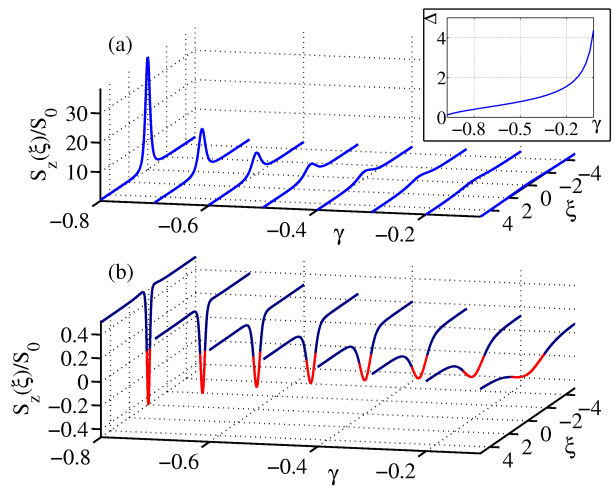}
\caption{Profile of the normalized Poynting vector $z-$component $S_z / S_0$ evaluated for the nonlinear guided waves of Fig.(3). Here
$\textrm{sign}(\epsilon) = -1$, $\textrm{sign}(\mu) = 1$ and $\textrm{sign}(\chi) = 1$. Panel (a) contains $S_z / S_0$ evaluated for the waves
reported in panel (a) and (b) of Fig.(3). Panel (b) contains $S_z / S_0$ evaluated for the waves reported in panel (c) and (d) of Fig.(3). In the
inset of panel (a), the width $\Delta$ (root-mean-square deviation) of the various $S_z / S_0$ reported in panel (a) is plotted as a function of
$\gamma$.}
\end{figure}
In Fig.(2) we plot various profiles of $u_x(\xi)$ and $u_z(\xi)$ for $\gamma = - 1/2$ and for $u_{x \infty}^2 < 1/2$. In Fig.(3) we plot the profiles
of $u_x(\xi)$ and $u_z(\xi)$ for different nonlinear guided waves, in the range $-0.8 < \gamma < -0.1$, each characterized by the same asymptotical
value $u_{x \infty} = \sqrt{0.4}$. As explained in Appendix A, more than one nonlinear guided wave (for a given $\gamma$) can be generally found for
each $u_{x \infty}$ and, in the situation of Fig.(3), there are specifically two waves, the first being reported in panel (a) and (b) and the second
in panel (c) and (d). The most striking feature emerging from Fig.(3) is that, for a given $u_{x \infty}$ the closer $\gamma$ to $-1$, the sharper
the profile of $u_x$, i.e. $\gamma$ produces an hyper-focusing effect for this waves when it approaches the value $\gamma = -1$.  As expected (see
the discussion of Sec.\ref{design}) the situation $\gamma = -1$ is peculiar displaying a phenomenology absent for other values of $\gamma$. From a
physical point of view the hyper-focusing effect results from the combination of two different mechanisms. For the first one, we note that, since
$\epsilon_{z}^{(NL)} (+\infty) = 0$ (see the second of Eqs.(\ref{efflim})), if $\gamma$ is very close to $-1$, from the second of Eqs.(\ref{eeff})
and the second of Eqs.(\ref{Ga1}), we conclude that $u_{z \infty}$ is much greater than one (in agreement with the second of Eqs.(\ref{beta})
evaluated for $\gamma \simeq -1$).  Therefore, $|u_z(\xi)|$ is much greater than one everywhere apart a small region around $\xi = 0$ where $u_z$ is
very small (since $u_z(0)=0$) and $du_z / d\xi$ is very large. The second mechanism supporting hyper-focusing is based on the fact that, for $\gamma
\simeq -1$ the field $x-$component ($u_x$) is nonlinearly driven solely by $u_z$ (see the discussion of Sec.\ref{design}). Therefore, in the region
where $u_z$ is very small and $du_z / d\xi$ is very large, from the first of Eqs.(\ref{syste}), it is evident that $u_x(\xi)$ displays a very
pronounced peak, i.e. $u_x$ is tightly squeezed by $u_z$ for $\gamma \simeq -1$. In Fig.(4) we report the $z$-component of the normalized Poynting
vector $S_z / S_0$ (see Eqs.(\ref{SSz1})) evaluated for the waves reported in Fig.(3). Note that the above discussed hyper-focusing effectively
corresponds to a tight energy localization around $\xi = 0$ for $\gamma$ close to $-1$. This effect is particularly evident from the inset of panel
(a) of Fig.(4) where we plot the peak width $\Delta$ of $S_z / S_0$ as a function of $\gamma$. The power flows reported in panel (b) of Fig.(4)
clearly displays the transverse power flow reversing discussed in Sec.\ref{PFR} since there is a region around $\xi =0$ where $S_z / S_0<0$ whereas
$S_z / S_0 > 0$ elsewhere. Note that the power flows reported in panel (a) of Fig.(4) does not exhibit transverse power flow reversing and this can
be easily understood considering the structures of the two waves reported in Fig.(3). The $z-$component of the first wave is such that $du_z/d\xi <
0$ (see panel (b) of Fig.(3)) so that, from the first of Eq.(\ref{SSz1}), the two bell shaped contributions $u_x^2$ and $-u_x d u_z / d\xi$ are both
positive yielding the $S_z / S_0 >0$. On the other hand, the second wave is such that $du_z/d\xi > 0$ (see panel (d) of Fig.(3)) and therefore, in
the first of Eq.(\ref{SSz1}), the two contributions have different signs and $S_z / S_0$ can flip its sign along the transverse profile.
\subsection{Nonlinear guided waves for $\sigma_\epsilon = -1$ and $\sigma_\mu =- 1$} \label{NGW2}
For $\sigma_\epsilon = -1$ and $\sigma_\mu =- 1$, waves are allowed to propagate in the linear regime and the nonlinear guided waves can propagate if
\begin{equation}
\beta^2 = 1 - \Psi_x (+\infty) > 0,
\end{equation}
(see the first of Eqs.(\ref{beta})), i.e. the background nonlinear plane wave can not produce a nonlinear dielectric response overcoming the linear
part.
\begin{figure}
\includegraphics[width=0.45\textwidth]{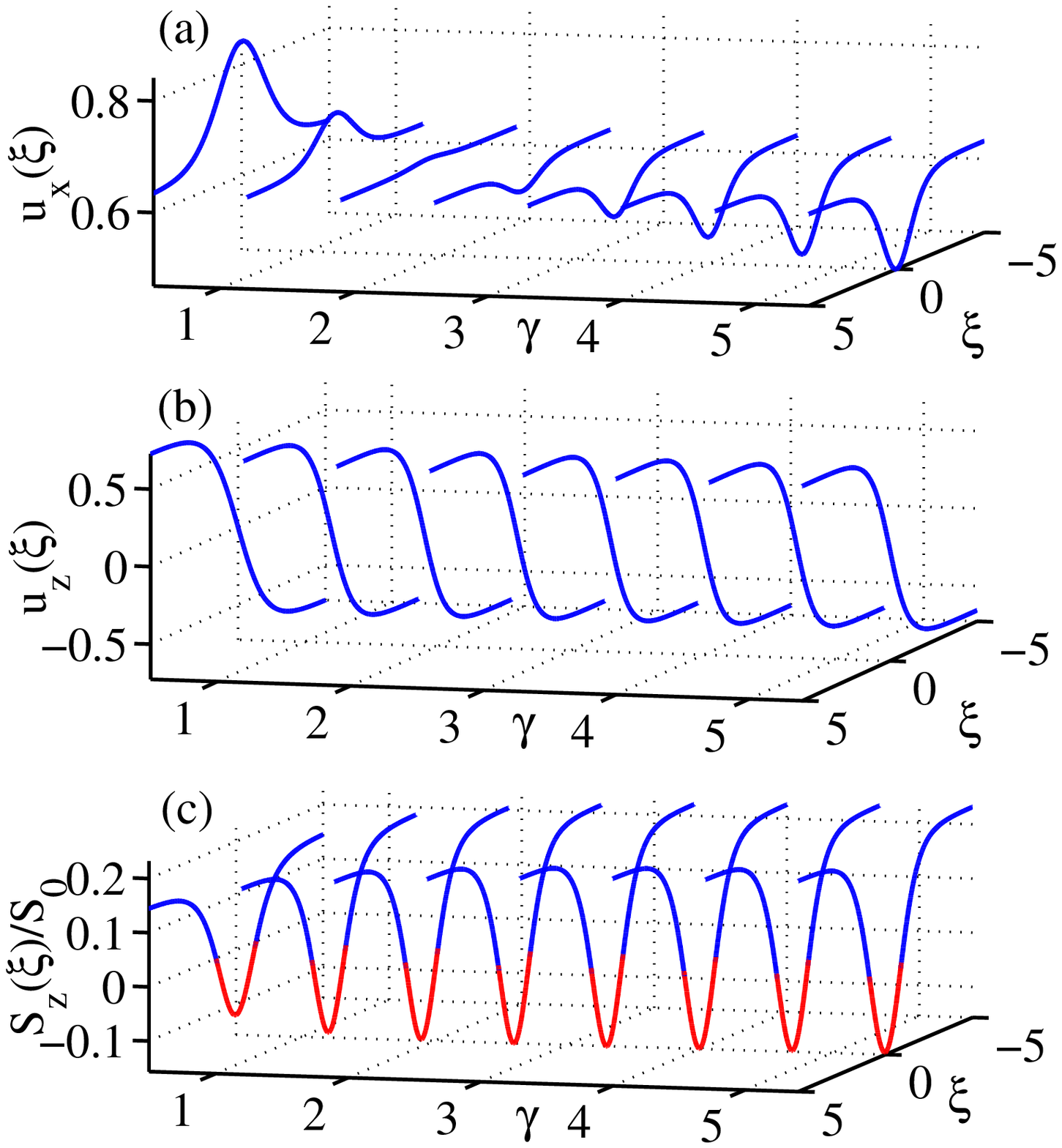}
\caption{Profiles of $u_x$ (a), $u_z$ (b) and $S_z/S_0$ (c) of various nonlinear guided waves with $\sigma_\epsilon=-1$ and $\sigma_\mu=-1$ and
$\gamma>0$. All the waves are characterized by the same asymptotical value $u_{x\infty} =\sqrt{0.4}$.}
\end{figure}
\begin{figure}
\includegraphics[width=0.45\textwidth]{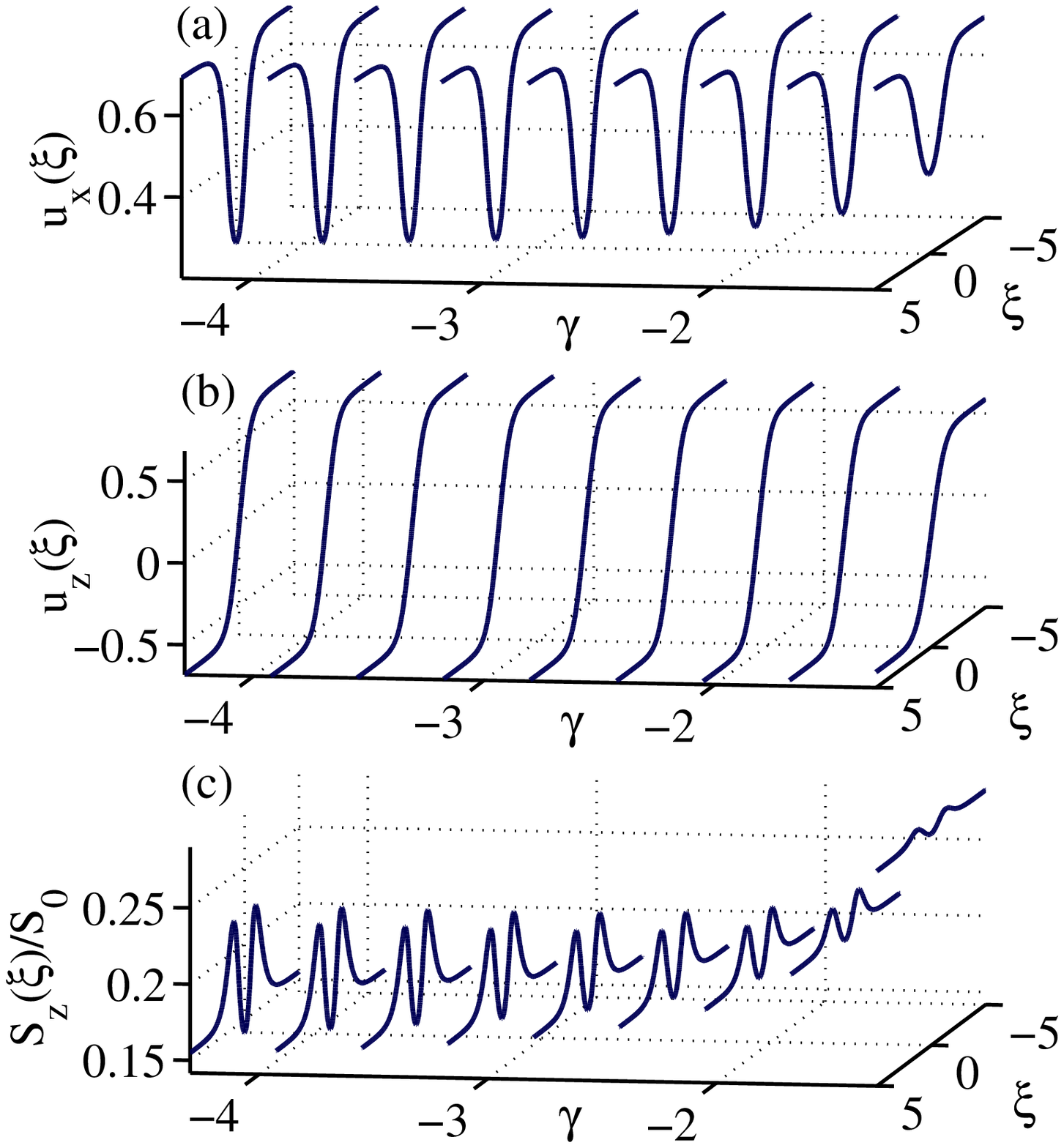}
\caption{Profiles of $u_x$ (a), $u_z$ (b) and $S_z/S_0$ (c) of various nonlinear guided waves with $\sigma_\epsilon=-1$ and $\sigma_\mu=-1$ and
$\gamma<-1$. All the waves are characterized by the same asymptotical value $u_{x\infty} =\sqrt{0.4}$.}
\end{figure}
As opposed to the case discussed in Sec.\ref{NGW1}, in the present situation there are two distinct families of nonlinear guided waves corresponding
to the ranges $\gamma>0$ and $\gamma<-1$ (see the second column of Table I). In Fig.5 we report different nonlinear guided waves profiles together
with their power flows for different values of $\gamma >0$ and for $u_{x \infty} = \sqrt{0.4}$. From panel (a) of Fig.5 we note that, in this regime,
the profile of $u_x$ has a peak and a hole (around $\xi = 0$) for small and large values of $\gamma$, respectively. This behavior is easily
understood since for $\gamma=0$ we have $\beta =0$ (from the first of Eqs.(\ref{beta})) and $u_{x}(0) = 1 > u_{x \infty}$ (from the first of
Eqs.(\ref{syste}) evaluated at $\xi = 0$ and for $\sigma_\epsilon = -1$) whereas, for $\gamma \rightarrow + \infty$ we have $\beta \rightarrow
\sqrt{2(1-2u_{x \infty}^2)}$ and $u_{x}(0) = 0 < u_{x \infty}$ (from the limit $\gamma \rightarrow +\infty$ of the first of Eqs.(\ref{syste}) for
$\sigma_\epsilon = -1$). As a consequence there must be a value of $\gamma$ for which $u_x$ is uniform. Requiring that $u_x(\xi) = u_{x\infty}$,
Eqs.(\ref{syste}) can be casted in the form
\begin{eqnarray} \label{syste1}
&& \beta \frac{d u_z}{d \xi} = (1-\gamma) \left(-u_{z \infty}^2 + u_z^2 \right) u_{x\infty} , \nonumber \\
&& \frac{d^2 u_z}{d \xi^2} = (1+\gamma)  \left(-u_{z \infty}^2 + u_z^2 \right) u_z
\end{eqnarray}
so that the solution of the second of Eqs.(\ref{syste1}) fulfilling the boundary conditions of Eqs.(\ref{boundary1}) is
\begin{equation} \label{dark}
u_{z}(\xi)=u_{z \infty} \textrm{tanh} \left[  \sqrt{\frac{1+\gamma}{2}} u_{z \infty} \xi \right]
\end{equation}
or, in other words, the longitudinal component $u_z$ is an exact electromagnetic dark soliton. However, the obtained $u_z$ has to satisfy the full
Maxwell system so that, substituting the field in Eq.(\ref{dark}) into the first of Eqs.(\ref{syste1}), it is straightforward to prove that this is
possible only if
\begin{equation} \label{gamma}
\gamma_\pm = \frac{1}{2 u_{x \infty}^2}\left[1 \pm \sqrt{1-4u_{x_\infty}^4} \right].
\end{equation}
For $u_{x \infty} = \sqrt{0.4}$ we obtain $\gamma_+ = 2$ which is the value of $\gamma$ at which $u_x$ is uniform in panel (a) of Fig.5. Note that
the situation where $u_x$ is uniform and $u_z$ is a dark soliton is possible only in the extreme nonlinear regime where the nonlinearity $\Psi_z$ can
compensate the linear part in the second of Eqs.(\ref{syste}). From panel (c) of Fig.5 it is evident that every considered waves exhibit the
transverse power flow reversing discussed in Sec.\ref{PFR}.

In Fig.6 we report different nonlinear guided waves profiles together with their power flows for different values of $\gamma <0$ and for $u_{x
\infty} = \sqrt{0.4}$. Let us consider, in this case, the behavior of the nonlinear guided waves for large values of $|\gamma|$. From panel (a) and
(b) of Fig.6 it is evident that a kind of saturation occurs, i.e. $u_x$ and $u_z$ approaches their asymptotic profiles for $\gamma \rightarrow
-\infty$. This is consistent with the fact that, taking the limit $\gamma \rightarrow -\infty$ of Eqs.(\ref{beta}) we obtain
\begin{eqnarray}
\beta^2 &=& 2\left(1-2u_{x \infty}^2\right) \nonumber \\
u_{z \infty}^2 &=&  u_{x \infty}^2
\end{eqnarray}
i.e. $\beta^2$ and $u_{z\infty}^2$ asymptotically approaches two finite asymptotic values and it is relevant that $u_{z \infty}^2 \rightarrow u_{x
\infty}^2$. In order to obtain the asymptotic profiles of $u_x$ and $u_z$, we note that, taking the limit $\gamma \rightarrow -\infty$ of
Eqs.(\ref{syste}) and consistently assuming that the profiles remains everywere finite, consistency requires that $u_{x}^2(\xi) = u_{z}^2(\xi)$ so
that, asymptotically, $u_{x}(\xi)=u_{z}(\xi)$ for $\xi <0$ and $u_{x}(\xi)=-u_{z}(\xi)$ for $\xi >0$ (since $u_x$ and $u_z$ are spatially even and
odd, respectively). Exploiting this property, Eqs.(\ref{syste}) yields, for $\gamma \rightarrow -\infty$, the equation
\begin{figure}
\includegraphics[width=0.45\textwidth]{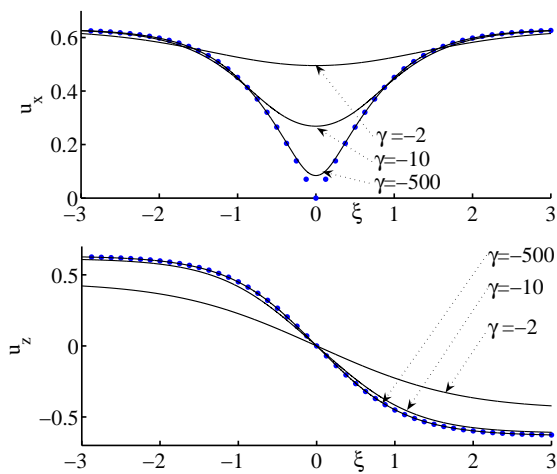}
\caption{Profiles of $u_x$ and $u_z$ (solid lines) of nonlinear guided waves for $u_{x\infty}=\sqrt{0.4}$ and $\gamma=-2,-10,-500$. The asymptotical
($\gamma \rightarrow -\infty$) profiles of Eqs.(\ref{asympt}) are also reported (dotted line).}
\end{figure}
\begin{equation}
\frac{d^2 u_z}{d \xi^2} = 4 \left(-u_{x \infty}^2 + u_z^2 \right) u_z
\end{equation}
so that, the asymptotic nonlinear guided wave profiles are
\begin{eqnarray} \label{asympt}
u_{x}(\xi) &=& \left| u_{x \infty} \textrm{tanh} \left(\sqrt{2} u_{x \infty} \xi \right) \right|, \nonumber \\
u_{z}(\xi) &=& -u_{x \infty} \textrm{tanh} \left(\sqrt{2} u_{x \infty} \xi \right).
\end{eqnarray}
Note that the asymptotic profile of $u_x$ is singular at $\xi = 0$ (i.e. it continuous but not differentiable) as a consequence of the unrealistic
assumption $\gamma \rightarrow -\infty$. Even though the profiles of Eqs.(\ref{asympt}) are not physical, they show that both $u_x^2$ and $u_z^2$
asymptotically approaches the square of an electromagnetic dark solitons and they provide a very good description of the above mentioned saturation
for $\gamma<0$. The profiles of $u_x$ and $u_z$ for $u_{x\infty}=\sqrt{0.4}$ and $\gamma=-2,-10,-500$ together with the asymptotical profiles of
Eqs.(\ref{asympt}) are plotted in Fig.(7) from which it is evident that, for $\gamma <-10$, the asymptotical profiles of Eqs.(\ref{asympt})
accurately describe the actual field profiles (apart a small region around $\xi = 0$ for $u_x$). We conclude that, in the extreme nonlinear regime,
the compensation between $\Psi_x$ and $\Psi_z$ and the linear terms, if $|\gamma|>>1$, forces $u_x^2$ and $u_z^2$ almost to coincide.
\section{Conclusions}
In conclusion, we have investigated the effective response of a nonlinear Kerr metamaterial obtained by homogenizing a one dimensional layered
periodic structure. The effective response formally coincides with that of a standard nonlinear Kerr medium with the important difference that its
parameters (both linear and nonlinear) can be independently designed and that they can assume even values not achievable in standard material. As a
consequence we can choose the linear dielectric permittivity to be much smaller than one thus allowing the observation of the regime where the
nonlinear polarization can not be regarded as a small perturbation (extreme nonlinear regime). As a leading general phenomenon characterizing the
extreme nonlinear regime we have discussed the transverse power flow reversing effect, i.e. the fact that the power flow can change its sign along
the transverse beam profile. Combining the extreme nonlinear regime and the fact the effective Kerr response can be tailored in an unconventional
way, we have discussed a number of novel phenomena exploiting a class of fields (nonlinear guided waves) admitting full analytical description.
Examples of such novel effects are the fact that the nonlinear waves can propagate even if the medium linear properties (dielectric permittivity and
magnetic permeability) would forbid propagation, hyper-focusing induced by the phase difference between the field components (in the case $\gamma =
-1$) and extreme compensation between the field components if they are $\pi /2$ out of phase in the limiting situation $\gamma >>1$.
\appendix
\section{Nonlinear guided waves existence}
In order to derive the existence conditions of nonlinear waves satisfying Eqs.(\ref{syste}) and Eqs.(\ref{boundary1}), it is convenient to cast
Eqs.(\ref{syste}) into the standard form of a first order system of differential equations. Differentiating the first of Eqs.(\ref{syste}) and
substituting the obtained expression of $d^2 u_z / d \xi^2$ into the second of Eqs.(\ref{syste}), we obtain, after some algebra
\begin{widetext}
\begin{eqnarray} \label{syst}
\beta \frac{d u_z}{d \xi} &=& (\beta^2- \sigma_\epsilon \sigma_\mu) u_x  - \sigma_\mu \left[(1+\gamma) u_x^2+(1-\gamma) u_z^2 \right] u_x, \nonumber \\
\beta \frac{d u_x}{d \xi} &=& \frac {\beta^2 \sigma_\epsilon +(1-\gamma)\left\{ (2\sigma_\epsilon \sigma_\mu - \beta ^2) + 2 \sigma_\mu \left[
(1+\gamma) u_x^2 + (1-\gamma) u_z^2 \right] \right\} u_x^2 + \beta^2 (1+\gamma) u_z^2}{\sigma_\epsilon + \left[ 3(1+\gamma) u_x^2 + (1-\gamma) u_z^2
\right]} u_z
\end{eqnarray}
\end{widetext}
which is a system of ordinary differential equations equivalent to Maxwell equations provided the relation
\begin{equation} \label{excond}
\sigma_\epsilon + \left[ 3(1+\gamma) u_x^2 + (1-\gamma) u_z^2 \right] \neq 0
\end{equation}
holds along the whole profile $u_x(\xi)$, $u_z(\xi)$ \cite{Ciatto}. The system of Eqs.(\ref{syst}) can be fully analytically investigated since it
admits the first integral
\begin{eqnarray} \label{F}
&& F(u_x,u_z) = \left[(\beta^2 - \sigma_\epsilon \sigma_\mu) u_x^2 - \sigma_\epsilon \sigma_\mu u_z^2 \right] \nonumber \\
           && -\frac{1}{2}  \sigma_\mu (1+\gamma) (u_x^4+u_z^4)- \sigma_\mu (1-\gamma) u_x^2 u_z^2  \nonumber \\
           && -\frac{1}{\beta^2}  \left\{ (\beta^2 - \sigma_\epsilon \sigma_\mu) - \sigma_\mu \left[(1+\gamma)u_x^2+(1-\gamma)u_z^2 \right] \right\}^2 u_x^2
               \nonumber \\
\end{eqnarray}
or, in other words, the relation
\begin{equation}
\frac{d}{d \xi} F(u_x(\xi),u_z(\xi)) =0
\end{equation}
holds for any solution $u_x(\xi),u_z(\xi)$ of Eqs.(\ref{syst}). Considering the boundary conditions of Eqs.(\ref{boundary1}), since $u_x(\xi)$ and
$u_z(\xi)$ have to asymptotically approach two constant values, their first and second derivatives vanish for $\xi \rightarrow +\infty$ so that we
require the right hand sides of Eqs.(\ref{syst}) to vanish at $u_x=u_{x\infty}$ and $u_z=u_{z\infty}$. As a consequence we obtain
\begin{eqnarray} \label{bet}
\beta^2 =\frac{2 \gamma \sigma_\mu }{1+\gamma} (\sigma_\epsilon  + 2  u_{x \infty}^2), \nonumber \\
u_{z \infty}^2 = \frac{-\sigma_\epsilon - (1-\gamma) u_{x \infty}^2}{1+\gamma},
\end{eqnarray}
which are the relations expressing the propagation constant $\beta$ and the asymptotical amplitude $u_{z \infty}$ as functions of the asymptotical
amplitude $u_{x \infty}$. In addition, exploiting the fact that the $F(u_x, u_z)$ is constant along the wave profile, the relation
\begin{equation} \label{cubic}
F(u_{x0},0) = F(u_{x\infty},u_{z\infty}),
\end{equation}
in which $\beta$ and $u_{z\infty}$ have been eliminated using Eqs.(\ref{bet}), is a cubic equation for $u_{x0}^2$ which can be solved to yield the
peak amplitude $u_{x0}$ as a function of the asymptotical amplitude $u_{x \infty}$.

For each $u_{x \infty}$, the existence of the corresponding guided waves has to be assured by a number of requirements. In first place $\beta$ and
$u_{z \infty}$ have to be real so that, from Eqs.(\ref{bet}), we obtain
\begin{eqnarray} \label{ineq1}
\frac{\gamma \sigma_\mu (\sigma_\epsilon  + 2  u_{x \infty}^2)}{1+\gamma} \geq 0, \nonumber \\
\frac{-\sigma_\epsilon - (1-\gamma) u_{x \infty}^2}{1+\gamma} \geq 0,
\end{eqnarray}
which are necessary inequalities for guided waves existence. Analogously, it is necessary that $u_{x0}$, obtained by Eq.(\ref{cubic}), is real. On
the other hand, the above boundary conditions for $\xi \rightarrow +\infty$ imply that, after substituting the expression of $\beta^2$ of the first
of Eqs.(\ref{bet}) into Eq.(\ref{F}), the function $F(u_x,u_z)$ has a stationary point at $(u_{x\infty},u_{z\infty})$. Therefore, since the curve
$(u_x(\xi),u_z(\xi))$ of the plane $(u_x,u_z)$ has to reach the point $(u_{x\infty},u_{z\infty})$, it has to be required that $F(u_x,u_z)$ has a
saddle point at $(u_{x\infty},u_{z\infty})$ and this leads to the necessary inequality
\begin{eqnarray} \label{ineq2}
\left[ \left( \frac{\gamma^2+4\gamma+1}{1+\gamma} \right) u_{x \infty}^2 + \frac{\gamma \sigma_\epsilon}{1+\gamma} \right] \nonumber \\
\times \left[2 (1-\gamma) u_{x \infty}^4 + \sigma_\epsilon (3-\gamma) u_{x \infty}^2 + 1 \right] >0.
\end{eqnarray}
We conclude that the existence of the nonlinear guided waves we are considering is assured by the condition of Eq.(\ref{excond}), together with the
reality of $u_{x0}$ (obtained by solving Eq.(\ref{cubic})) and the inequalities of Eqs.(\ref{ineq1}) and (\ref{ineq2}). The fulfilment of all these
requirements leads to the existence conditions reported in Table I.

%%%%%%%%%%%%%%%%%%%%%%%%%%%%%%%%%%%%%%%%%%%%%%%%%%%%%%%%%%%%%%%%%%%%%%%%%%%%%%%%%%

\end{document}